\documentclass[runningheads,orivec]{llncs} 

\usepackage[T1]{fontenc}
\usepackage{graphicx,verbatim}
\usepackage{booktabs}

\usepackage{amsmath}
\usepackage{amssymb}
\usepackage{mathtools}
\usepackage{xcolor}
\definecolor{firebrick}{RGB}{178,34,34}
\xdefinecolor{jhublue}{HTML}{002D72} 
\xdefinecolor{jhubluemedium}{HTML}{005eb8} 
\xdefinecolor{jhubluelight}{HTML}{68ACE5} 

\usepackage[
  colorlinks=true,
  linkcolor=firebrick,
  citecolor=firebrick,
  urlcolor=firebrick,
  pdfborder={0 0 0}
]{hyperref}

\usepackage[nameinlink,noabbrev]{cleveref}
\crefname{figure}{Fig.}{Figs.}
\crefname{table}{Tab.}{Tabs.}
\crefname{section}{Sec.}{Secs.}
\Crefname{section}{Sec.}{Secs.}

\usepackage{bbm}
\usepackage{physics}

\usepackage{tikz}

\newcommand{\calL}{\mathcal{L}}

\newcommand{\bbx}{\mathbf{x}}

\usepackage[normalem]{ulem}

\begin{document}
\title{HemoPIC: A Physics-Informed Cerebral Hemodynamics Digital Twin for Brain Perfusion}
\titlerunning{HemoPIC}
%

\author{Yi-Chen Lee \and Peirong Liu\thanks{Corresponding author.}}  
\authorrunning{Y. Lee and P. Liu}
\institute{Department of Electrical and Computer Engineering,\\
Data Science and AI Institute,\\
Johns Hopkins University\\ \vspace{0.15cm}
\email{\{ylee268,pliu53\}@jh.edu}
}
  
\maketitle              

\sloppy 

\begin{abstract}
Perfusion imaging guides clinical evaluation of stroke and brain tumors by characterizing tissue-level hemodynamics. Routine quantification relies on manual arterial input function (AIF) selection followed by deconvolution, producing summary maps without an executable temporal model for simulation or mechanistic insight. Tracer-dynamics-based models infer transport or compartmental parameters from perfusion time series, but do not yield clinically actionable perfusion indices (e.g., CBF, CBV, MTT) that inform diagnosis and treatment decisions. In this work, we propose HemoPIC, a physics-informed cerebral hemodynamics digital twin that explains perfusion time series through tracer mass conservation and a lumped parameter hemodynamic model. Specifically, HemoPIC solves a constrained inverse problem that jointly estimates digital twin parameters and latent states from perfusion imaging, eliminating manual AIF selection and deconvolution from routine perfusion quantification while directly producing clinically actionable perfusion summary maps. Experiments demonstrate that HemoPIC reconstructs tracer dynamics, generates physiologically consistent perfusion maps with lesion hypoperfusion patterns, satisfies central volume consistency, and yields a mechanistic hemodynamic digital twin that enables forward simulation and counterfactual intervention analysis. Code is publicly available at \url{https://github.com/jhuldr/HemoPIC}.


\end{abstract}

\section{Introduction}
\label{sec:intro}

Medical digital twins construct patient-specific models that link clinical observations to latent physiological states. A standard twin supports patient monitoring, forecasting, and risk stratification. A physics-informed twin goes further. It encodes conservation laws and physiologic dynamics, enabling forward simulations, backward inference to reveal the physiology underlying clinical observations, and counterfactual evaluation of hypothetical interventions. 
Although digital twins have been explored for organs such as the heart, lung, and liver~\cite{katsoulakis2024digital,kuang2024med}, cerebral digital twins remain underexplored, largely due to the brain's structural complexity and the
identifiability bottleneck under routine clinical measurements. In particular, variability in the Circle of Willis and collateral flow pathways complicates reliable inference of regional perfusion under occlusion~\cite{hoksbergen2000collateral}. 

Perfusion imaging, such as dynamic susceptibility contrast (DSC) MRI which measures cerebral blood flow via intravascular tracer injections, is widely used to assess cerebral perfusion and vascular permeability in stroke, brain tumors, and neurodegeneration~\cite{desai2007blood,kermode1990breakdown}.
Clinically, perfusion image analysis typically derives spatial summary maps, such as cerebral blood flow (CBF), cerebral blood volume (CBV), and mean transit time (MTT), from the observed tracer time series using voxelwise kinetic modeling~\cite{fieselmann2011deconvolution}. 
However, this process requires manual selection of an arterial input function (AIF) followed by deconvolution, a sensitive inverse approximation step that could introduce substantial variability in the resulting perfusion summary maps~\cite{kudo2010differences}. 
Because treatment decisions (e.g., thrombectomy) rely on static thresholding of these maps, small algorithmic differences can lead to large discrepancies in penumbra estimation and potentially inconsistent clinical decisions. 
More fundamentally, these summary maps do not define an explicit hemodynamic state evolution under a governing physical law, limiting their use for patient-specific modeling and simulation~\cite{liu2021perfusion}. 

Several recent efforts move beyond static map-based postprocessing by incorporating physics-informed modeling. Tracer transport approaches~\cite{liu2020piano,liu2021perfusion,liu2021discovering} fit an advection-diffusion model, governed by spatially varying blood velocity and diffusion fields, directly to the spatiotemporal tracer dynamics. These approaches enforce a physical law and capture spatial coupling beyond voxelwise kinetic fitting.
Quantitative transport mapping~\cite{zhou2021quantitative} applies a conservation law tracer transport model to infer flow-related fields from dynamic imaging, demonstrated in the kidney with multi-delay arterial spin labeling and angiography-informed vascular priors, and in the liver with dynamic contrast-enhanced (DCE) MRI~\cite{romano2025validation,zhang2022fluid}.
These transport-derived fields can be estimated without manual AIF selection.
However, translating them into clinically actionable perfusion indices such as CBF, remains nontrivial.
One notable exception is \cite{briggs2025physics}, which formulates a physics-informed cerebral hemodynamics digital twin with latent states defined directly in volumetric CBF; however, it requires multimodal bedside monitoring signals, including arterial blood pressure (ABP), intracranial pressure (ICP), transcranial Doppler \cite{claassen2007transcranial} and end tidal carbon dioxide \cite{Levitzky2018Pulmonary} other than perfusion imaging,  which are not standardized in routine clinical practice.

\noindent \textbf{Contribution:}
In this work, we introduce HemoPIC:
a \underline{P}hysics-\underline{I}nformed \underline{C}erebral \underline{Hemo}dynamics digital twin constructed from routine perfusion imaging alone. HemoPIC infers temporal hemodynamic latent states that govern the full tracer inflow-outflow passage and enable forward simulation. By introducing a lumped terminal hemodynamic closure on spatially aggregated tracer dynamics, HemoPIC links blood inflow, cerebrovascular storage, and blood outflow within a compact physical structure without patient-specific outlet conditions. Notably, HemoPIC removes manual AIF selection, bypasses deconvolution pipelines, and yields hemodynamic latent state estimates that directly lead to clinically interpretable perfusion summary maps. 
Experiments show that HemoPIC generates clinically consistent perfusion summary maps, satisfies central volume theorem consistency, and reveals underlying Windkessel dynamics. Importantly, it is the first method to reconstruct the complete inflow–outflow tracer passage.

\section{Patient-Specific Hemodynamic Digital Twin for Cerebral Perfusion via Tracer Transport}
\label{sec:method}

\begin{figure}[t]
\centering
\resizebox{1.\textwidth}{!}{
	\begin{tikzpicture} 

    \tikzstyle{mylines}=[line width=0.8mm]
    \pgfmathsetmacro{\sx}{2.5}
    \pgfmathsetmacro{\dx}{1.5}
    \pgfmathsetmacro{\sy}{5}
    \pgfmathsetmacro{\dy}{1}

	\node[right] at (\sx - 1.6, \sy - \dy) {\small $C(t)$:}; 
	\node at (\sx, \sy - \dy) {\includegraphics[width=0.11\textwidth]{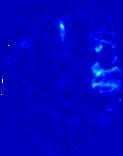}};  
	\node at (\sx + \dx, \sy - \dy) {\includegraphics[width=0.11\textwidth]{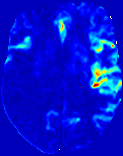}};
	\node at (\sx + 2*\dx, \sy - \dy) {\includegraphics[width=0.11\textwidth]{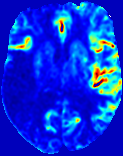}};
	\node at (\sx + 3*\dx, \sy - \dy) {\includegraphics[width=0.11\textwidth]{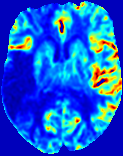}};
	\node at (\sx + 4*\dx, \sy - \dy) {\includegraphics[width=0.11\textwidth]{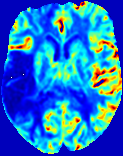}};
	\node at (\sx + 5*\dx, \sy - \dy) {\includegraphics[width=0.11\textwidth]{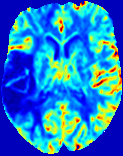}};
	\node at (\sx + 6*\dx, \sy - \dy) {\includegraphics[width=0.11\textwidth]{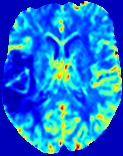}};
	\node at (\sx + 7*\dx, \sy - \dy) {\includegraphics[width=0.11\textwidth]{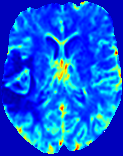}};
	\node at (\sx + 8*\dx, \sy - \dy) {\includegraphics[width=0.11\textwidth]{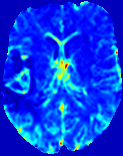}};
	\node at (\sx + 9*\dx, \sy - \dy) {\includegraphics[width=0.11\textwidth]{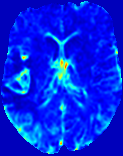}};

	\draw[color = jhubluelight!20, line width=0.6mm,rounded corners=3, fill = jhubluelight!20] (\sx - 2.35, \sy - 2.4) rectangle +(18., -3.5);  
	\node[rotate=90] at (\sx - 2.7, \sy-\dy) {\footnotesize \textbf{Measured}};  
	\node[rotate=90] at (\sx - 2.7, \sy-4.25) {\footnotesize \textbf{Underlying}}; 
	\draw[->, line width=0.6mm, color = gray] (\sx - 1.6, \sy - \dy) -- (\sx - 2, \sy - \dy) -- (\sx - 2, \sy - 2.9)-- (\sx - 1, \sy - 2.9);
	\draw[-, line width=0.6mm, color = gray] (\sx - 2, \sy - 2.9) -- (\sx - 2, \sy - 3.5 - 0.8) -- (\sx - 1.6, \sy - 3.5 - 0.8);
	\draw[->, line width=0.6mm, color = gray] (\sx + 14.5, \sy - 2.9) -- (\sx + 15.3, \sy - 2.9) -- (\sx + 15.3, \sy - \dy) -- (\sx + 14.5, \sy - \dy); 
    \node at (\sx + 6.75, \sy - 2.93) {\textbf{Tracer Transport~(Sec.~\ref{subsec:model}):} $f_{\mathrm{in}}^r(t) \coloneqq \int_ {\Gamma_{ \mathrm{in}}^r(t) } \big( -u(\bbx, t) \cdot n_r(\bbx) \big) \dd S, \quad 
f_{\mathrm{out}}^r(t) \coloneqq \int_{ \Gamma_{\mathrm{out}}^r(t) } \big( u(\bbx, t) \cdot n_r \big) \dd S$};

	\draw[color = gray, line width=0.6mm,rounded corners=3, fill=gray!20] (\sx - 1.6, \sy-3.5) rectangle +(11.6, -1.6); 

    \node at (\sx - 1.6 + 5.8, \sy - 3.5 - 0.5) {\textbf{Digital twin parameters:} $\Theta = \bigcup_{r=1}^N \Theta_r,$ obtained by minimizing $\calL$ in \eqref{eqn:fitting}};
    
    \node at (\sx - 1.6 + 5.8, \sy - 3.5 - 1.6 + 0.5) {through the tracer dynamics and lumped Windkessel coupling derived in \eqref{eqn:compartment}};
    
    \node at (\sx - 1.6 + 5.8, \sy - 5.5) {\textbf{Estimating Underlying Physical Dynamics Parameters~(Sec.~\ref{subsec:estimation})}};

    \pgfmathsetmacro{\sx}{0.8}
	\pgfmathsetmacro{\sy}{\sy - 4.5} 

	\draw[<->, line width=0.6mm, color = gray](\sx + 11.9, \sy + 0.2) -- (\sx + 12.5, \sy + 0.2); 

	\node at (\sx-1.65 + 15, \sy + 0.2) {\includegraphics[width=0.11\textwidth]{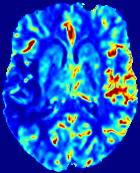}}; 
    \node at (\sx - 1.65 + 15, \sy - 1.) {{\footnotesize CBF}};
	\node at (\sx-1.65 + 15 + 1.5, \sy + 0.2) {\includegraphics[width=0.11\textwidth]{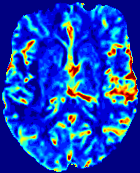}}; 
    \node at (\sx - 1.65 + 15 + 1.5, \sy - 1.) {{\footnotesize CBV}};
	\node at (\sx-1.65 + 15 + 3, \sy + 0.2) {\includegraphics[width=0.11\textwidth]{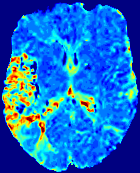}}; 
    \node at (\sx - 1.65 + 15 + 3, \sy - 1.) {{\footnotesize MTT}};

    \pgfmathsetmacro{\sx}{2.5}
	\pgfmathsetmacro{\sy}{\sy + 0.1} 
    
	\node at (\sx + 9.8*\dx, \sy + 2.75) {$t (s)$};
	\draw[mylines, color = black, line width=0.5mm, -] (\sx + 4.5*\dx, \sy+2.3) -- (\sx + 0*\dx, \sy+2.3) -- (\sx + 0*\dx, \sy+2.5);
	\draw[mylines, color = black, line width=0.5mm, -](\sx + 1*\dx, \sy+2.3) -- (\sx + 1*\dx, \sy+2.5);
	\draw[mylines, color = black, line width=0.5mm, -](\sx + 2*\dx, \sy+2.3) -- (\sx + 2*\dx, \sy+2.5);
	\draw[mylines, color = black, line width=0.5mm, -](\sx + 3*\dx, \sy+2.3) -- (\sx + 3*\dx, \sy+2.5);
	\draw[mylines, color = black, line width=0.5mm, -](\sx + 4*\dx, \sy+2.3) -- (\sx + 4*\dx, \sy+2.5);
	\draw[mylines, color = black, line width=0.5mm, -](\sx + 5*\dx, \sy+2.3) -- (\sx + 5*\dx, \sy+2.5);
	\draw[mylines, color = black, line width=0.5mm, -](\sx + 6*\dx, \sy+2.3) -- (\sx + 6*\dx, \sy+2.5);
	\draw[mylines, color = black, line width=0.5mm, -](\sx + 7*\dx, \sy+2.3) -- (\sx + 7*\dx, \sy+2.5);
	\draw[mylines, color = black, line width=0.5mm, -](\sx + 8*\dx, \sy+2.3) -- (\sx + 8*\dx, \sy+2.5);
	\draw[mylines, color = black, line width=0.5mm, -](\sx + 9*\dx, \sy+2.3) -- (\sx + 9*\dx, \sy+2.5);
	\draw[mylines, color = black, line width=0.5mm, -](\sx + 4.5*\dx, \sy+2.3) -- (\sx + 9*\dx, \sy+2.3) -- (\sx + 9*\dx, \sy+2.5);
	\draw[mylines, color = black, line width=0.5mm, ->](\sx + 9*\dx, \sy+2.3) -- (\sx + 9.8*\dx, \sy+2.3);

	\end{tikzpicture}
	}
  \vspace{-0.5cm}
  \caption{\textbf{HemoPIC overview.} Given measured tracer dynamics, HemoPIC estimates regional blood inflow ($f^r_{ \mathrm{in}}$) and outflow ($f^r_{ \mathrm{out}}$) without requiring AIF, by coupling tracer transport with patient-specific hemodynamics in a physics-grounded formulation.
  }
  \label{fig:framework}
\end{figure}
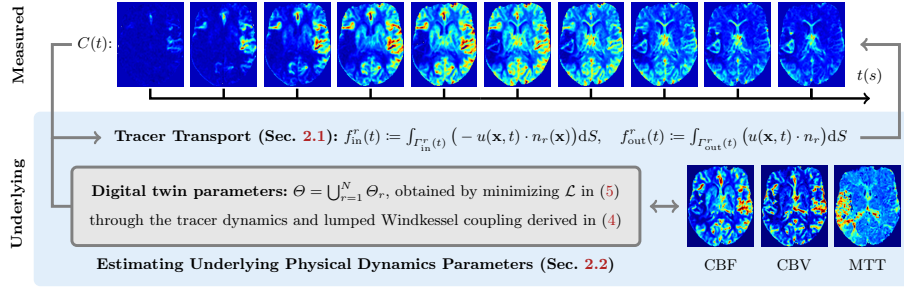

\cref{subsec:model} formulates the forward model that links perfusion tracer dynamics with a lumped Windkessel model. \cref{subsec:estimation} presents a parameter inference procedure by solving the resulting composite inverse problem from observed tracer dynamics. 
\cref{fig:framework} provides an overview of the HemoPIC framework.

\subsection{Regional Cerebral Inflow-Outflow Model Formulation} 
\label{subsec:model}


High-dimensional cerebrovascular simulations solve incompressible conservation laws to estimate intravascular blood flow velocity and pressure fields, but their predictions are highly sensitive to boundary conditions, with inlet waveforms and physiological assumptions significantly influencing hemodynamic outputs~\cite{mut2014effects,xiang2014effect}. Furthermore, intravascular fields alone do not directly yield clinical perfusion indices (e.g., CBF) without additional tracer kinetic modeling~\cite{fieselmann2011deconvolution,meier1954theory}, and outlet boundary conditions typically require terminal lumped resistance models~\cite{Korte2024MultiDimensional}.

To mitigate this boundary-condition sensitivity, 
we adopt a regionally aggregated formulation.
Specifically, we consider a partition of the brain domain $\Omega \subset \mathbb{R}^{3}$ into $N$ disjoint subsets $\{\Omega _{r}\}_{r=1}^{N}$, such that $\Omega =\bigcup _{r=1}^{N} \Omega _{r}$. Each subset $\Omega_{r}$ denotes a spatial cluster with boundary $\partial \Omega_{r}$ and an associated outward unit normal vector $n_r(\bbx)$ for $\bbx \in \partial \Omega_r$. 
These subsets correspond to subject-specific anatomical units, such as vascular compartments defined by tissue geometry and structural segmentation~\cite{billot2023synthseg}. 
Let $u(\bbx, t) \coloneqq (u^x(\bbx,t), u^y(\bbx,t), u^z(\bbx,t) )^\top$ denote the spatiotemporal blood flow velocity in directions $x, y, z$, respectively. For a given $\Omega_r$, we partition the boundary as $\partial\Omega_r=\Gamma_{\mathrm{in}}^r(t) \cup \Gamma_{\mathrm{out}}^r(t)$, where $\Gamma_{\mathrm{in}}^r(t)$ and $\Gamma_{\mathrm{out}}^r(t)$ denote the boundary of blood inflow and outflow, respectively,
$$\Gamma_{\mathrm{in}}^r(t) \coloneqq \big\{ \bbx \in \partial \Omega_r : u(\bbx,t) \cdot n_r < 0 \big\}, \quad
\Gamma_{\mathrm{out}}^r(t) \coloneqq \big\{\bbx \in \partial \Omega_r : u(\bbx, t) \cdot n_r > 0 \big\}.$$

We then define the effective volumetric inflow and outflow rates for $\Omega_r$ by
$$f_{\mathrm{in}}^r(t) \coloneqq \int_ {\Gamma_{ \mathrm{in}}^r(t) } \big( -u(\bbx, t) \cdot n_r(\bbx) \big) \dd S, \quad 
f_{\mathrm{out}}^r(t) \coloneqq \int_{ \Gamma_{\mathrm{out}}^r(t) } \big( u(\bbx, t) \cdot n_r \big) \dd S.$$


\paragraph{\textbf{Lumped Windkessel Model.}} 
To link regional inflow and outflow without subject-specific outlet boundary data, we incorporate a two-element Windkessel model as a terminal impedance boundary condition for each tracer compartment. Specifically, for each partition $\Omega_r$, the model couples the effective inflow and outflow rates $( f_{ \mathrm{in} }^r(t), f_{ \mathrm{out} }^r(t))$ through regional lumped parameters $\theta^r = (\theta_W^r, \theta_R^r)$, 
$$\theta_W^r \frac{ \dd }{ \dd t} ( P^r(t) - P_v^r ) = f_{ \mathrm{in} }^r(t) - f_{ \mathrm{out} }^r(t), \quad f_{ \mathrm{out} }^r(t) = \frac{ P^r(t) - P_v^r }{\theta_R^r},$$
where $P^r(t)$ denotes an effective microvascular pressure state, $P_v^r$ denotes a constant downstream reference pressure, $\theta_R^r>0$ denotes an effective resistance, and $\theta_W^r>0$ denotes an effective compliance. Equivalently, we have the expression:
\begin{align}
\label{eqn:windkessel}
f_{\mathrm{out}}^r(t) = f_{\mathrm{out}}^r(0) \exp(-\frac{t}{\tau^r}) + \frac{1}{\tau^r} \int_0^t \exp(\frac{s-t}{\tau^r}) f_\mathrm{in}^r(s) \dd s,
\end{align}
where $\tau^r = \theta_W^r \cdot \theta_R^r$.
Within each partition $\Omega_r$, $\theta_W^r$ represents vascular compliance (blood storage), $\theta_R^r$ represents resistance (blood discharge), and product $\tau^r$ sets the characteristic relaxation time governing outflow response to inflow changes.


\paragraph{\textbf{Tracer Dynamics.}}
We further link the above regional inflow-outflow formulation to perfusion tracer dynamics.
Let $C_r(\bbx, t)$ denote the tracer concentration  at $\bbx\in\Omega_r$, at time $t$. The advection-diffusion equation~\cite{liu2020piano,liu2021perfusion} gives:
\begin{align}
\label{eqn:advec_diffuse}
\frac{ \partial C_r(\bbx,t) }{ \partial t} = -\nabla \cdot ( u(\bbx, t) C_r(\bbx,t) ) + \nabla \cdot ( D(\bbx, t) \nabla C_r(\bbx,t) ),
\end{align}
where $D$ governs the tracer's diffusion. 
Integrating both sides over $\Omega_r$ and applying the divergence theorem, the right-hand side of \eqref{eqn:advec_diffuse} becomes:
\begin{align}
\label{eqn:pde_divergence}
\underbrace{ - \int_{\partial \Omega_r} ( u(\bbx, t) C_r(\bbx,t) ) \cdot n_r \dd S }_{ \coloneqq \Phi_A^r(t)} + \underbrace{ \int_{\partial \Omega_r}  ( D(\bbx,t) \nabla C_r(\bbx,t) ) \cdot n_r \dd S }_{ \coloneqq \Phi_D^r(t)},
\end{align} 
where $\Phi_A^r(t)$ admits the decomposition:
\begin{align*}
\Phi_A^r(t) = -  \int_{\Gamma_{ \mathrm{out} }^r(t)} ( u(\bbx, t) \cdot n_r ) C_r(\bbx,t)\dd S  - \int_{\Gamma_{ \mathrm{in} }^r(t) } ( u(\bbx, t) \cdot n_r ) C_r(\bbx,t) \dd S.
\end{align*}
To compactly represent the advective flux, we define the weighted concentration flux, with a minus sign for inflow and a plus sign for outflow:
$$C_{\mathrm{in/out}}^r(t) ~ \coloneqq 
\frac{1}{f_{\mathrm{in/out}}^r(t)} \cdot 
\int_{\Gamma_{\mathrm{in/out}}^r(t)}
\pm\,(u(\mathbf{x},t)\!\cdot\! n_r(\mathbf{x}))  C_r(\mathbf{x},t)\, \dd S, $$
For an effective distribution volume $V_r$, we define the cluster level state average concentration $\bar{C}_r(t) \coloneqq \frac{1}{V_r} \int_{\Omega_r} C_r(\bbx ,t) \dd \bbx$.
Then, \eqref{eqn:pde_divergence} becomes
\begin{align}
\label{eqn:compartment}
\frac{\dd}{\dd t}( V_r \bar{C}_r(t) ) = f_{ \mathrm{in} }^r (t) C_{\mathrm{in}}^r (t) - f_{ \mathrm{out} }^r (t)  C_{\mathrm{out}}^r (t) + \Phi_D^r(t).
\end{align}
We adopt a diffusion-free formulation with $\Phi_D^r(t)=0$ in \eqref{eqn:compartment} for all regions $r \in [N]$, considering diffusion is negligible compared to advection~\cite{liu2021perfusion}.

\subsection{Estimating Underlying Physical Dynamics Parameters}
\label{subsec:estimation}

Here, we formulate the estimation problem for the digital twin in \cref{subsec:model}, from perfusion-derived volumetric tracer time series. 
Since perfusion source data do not uniquely identify region-specific inflow rates and boundary conditions, we adopt a reduced parameterization that preserves the characteristic relaxation scale of the Windkessel, while keeping the inverse problem identifiable. 
Specifically, 
for each $r \in [N]$, we compute the regional mean concentration $\big\{ ( \bar{C}_r(t) \mid t \in [T]  \big\}$ by averaging voxelwise concentrations over time,
and consider a regional constant inflow $f_{ \mathrm{in} }^r(t) = F_r$. 
We represent the inflow concentration through a region-specific amplitude $\bar{C}_{\mathrm{in}}^r > 0$, modulating a shared inflow shape $C_{\mathrm{in}}^r(t) = \bar{C}_{\mathrm{in}}^r \cdot C_{\mathrm{art}}(t) $, where $C_{\mathrm{art}}(t)$ is a unit peak arterial proxy curve shared across clusters. The resulting composite model contains two parts. 
First, the lumped Windkessel model computes the blood outflow $\hat{f}_{\mathrm{out}}^r(t)$ from $(F_r, \theta_W^r, \theta_R^r)$ under a fixed initial condition \eqref{eqn:windkessel}. Second, the reduced tracer compartment yields the regional concentration $\hat{C}_r(t)$ with $(V_r, C_{ \mathrm{in} }^r(t), \hat{f}_{\mathrm{out}}^r(t))$. 
Together, we estimate the digital twin 
parameters,
$
\Theta = \bigcup_{r=1}^N \Theta_r, \text{~where~} 
\Theta_r = \big\{ \theta_W^r, \theta_R^r, V_r, F_r, \bar{C}_{\mathrm{in}}^r \big\},
$
by minimizing the following loss with regularization coefficients  $\lambda_{ \mathrm{in} }, \lambda_V, \lambda_F, \lambda_R$:
\begin{align}
\label{eqn:fitting}
\hspace*{-0.15cm}
\calL  = \sum_{r=1}^N  \frac{ \sum_{t=1}^T \omega(t) \big( \hat{C}_r(t; \Theta_r)  - \bar{C}_r(t) \big)^2}{\sum_{t=1}^T \omega(t) \bar{C}_r(t)^2 + \epsilon} + \lambda_{\mathrm{in}} \calL_{\mathrm{in}} + \lambda_V \calL_V + \lambda_F \calL_F + \lambda_R \calL_R,
\end{align}
subject to $\theta_R^r, \theta_W^r, F_r, V_r, \bar{C}_{\mathrm{in}}^r > 0$ for all $r$, where $\omega(\cdot)$ is a positive time weight that modulates each time point’s contribution, $\calL_{\mathrm{in}} = \sum_r ( \log{ \bar{C} }_{\mathrm{in}}^r - \log{C}^r_{\mathrm{prior}})^2$, $\calL_V = \sum_r ( \log{V}_r - \log{V}_r^{\mathrm{prior}})^2$ and $\calL_F = \sum_r ( \log{F}_r - \log{F}_r^{\mathrm{prior}} )^2$ with positive priors $\bar{C}_{\mathrm{prior}}^r, V_r^{\mathrm{prior}}$ and $F_r^{\mathrm{prior}}$ derived from data to stabilize the inverse problem. We additionally apply $\calL_R = \sum_r ( \log{\theta}_R^r - \log{\theta}_R^0 )^2$, where $\theta_R^0 > 0$ is a fixed reference resistance used to avoid extreme values under limited identifiability.

\section{Experiments}
\label{sec:exp}

We evaluate HemoPIC on the ISLES 2017 acute ischemic stroke dataset~\cite{winzeck2018isles}, consisting of 44 patients with 4D DSC-MRI perfusion images, deconvolution-derived perfusion maps (CBF, CBV, MTT), and gold-standard expert-annotated lesion masks.
MR signals are converted to voxelwise tracer concentration~\cite{fieselmann2011deconvolution}.

To obtain anatomically meaningful partitions $\{\Omega_r\}_{r=1}^N$, we apply SynthSeg~\cite{billot2023synthseg} to segment grey matter (GM), white matter (WM), cerebrospinal fluid (CSF), and brain stem. GM and WM are further divided using k-means on tracer time-curve features (8 clusters for GM, 6 for WM), neighboring clusters with fewer than 400 voxels are merged. 
For each $\Omega_r$, parameters are fitted from the regional mean concentration curve. A global arterial curve $C_{\mathrm{art}}(t)$ is constructed from high-peak time curves, smoothed (Gaussian std $=$ 1.8), and peak-normalized. 
$F_r^{\mathrm{prior}}$ ($V_r^{\mathrm{prior}}$) is set from median CBF (CBV) within $\Omega_r$, $\bar{C}_{\mathrm{prior}}^r$ is set from the peak of smoothed $C_r(t)$. 
We optimize using Powell’s method with $\lambda_{\mathrm{in}} = 0.06$, $\lambda_V = \lambda_F = 0.04$, $\lambda_R = 0.002$, $ \theta_R^0 = 1$. Loss weights $\omega(t)$ are set as $0.2 + 0.8 C_{\mathrm{art}}(t)$, restricting evaluation to time points with $\omega(t) \ge 0.2$ to emphasize tracer passage.


\begin{table}[ht]
\centering
\setlength{\tabcolsep}{3.pt}
\renewcommand{\arraystretch}{1.0}
\scriptsize
\caption{Cohort summary of normalized root mean square error for tracer concentration reconstruction ($N=44$). Q25 and Q75 denote the 25th and the 75th percentiles. 
}
\label{tab:ctc_nrmse_summary}
\label{table:ctc}
\resizebox{\linewidth}{!}{%
\begin{tabular}{@{}l c c c c | c@{}}
\toprule
Stats & GM & WM & Lesion & c-Lesion & Full brain \\
\midrule
mean $\pm$ std~ & ~0.106$\pm$0.045~ & ~0.095$\pm$0.068~ & ~0.318$\pm$0.291~ & ~0.192$\pm$0.104~ & ~0.093$\pm$0.057~ \\
median & 0.099 & 0.083 & 0.266 & 0.176 & 0.083 \\
{}[Q25, Q75] & [0.085,0.108] & [0.066,0.095] & [0.162,0.361] & [0.148,0.210] & [0.075,0.088] \\
\bottomrule
\end{tabular}%
}
\end{table}



\begin{figure}[t]

\centering 

\resizebox{1\linewidth}{!}{
	\begin{tikzpicture} 

        
	\pgfmathsetmacro{\shift}{-3.2}
	\pgfmathsetmacro{\dx}{0.3}


	\node at (5.2+\shift+0.1*3, 6.2+0.15*3, 0.75*3) {\scriptsize{Lesion}}; 
	\node at (5.2+\shift+0.1*3, 5+0.15*3, 0.75*3) {\includegraphics[width=0.16\textwidth]{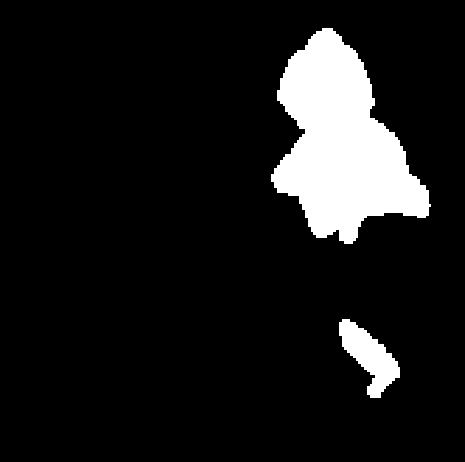}};
	\node at (5.20+\shift+0.1*3, 3+0.15*3, 0.75*3) {\includegraphics[width=0.16\textwidth]{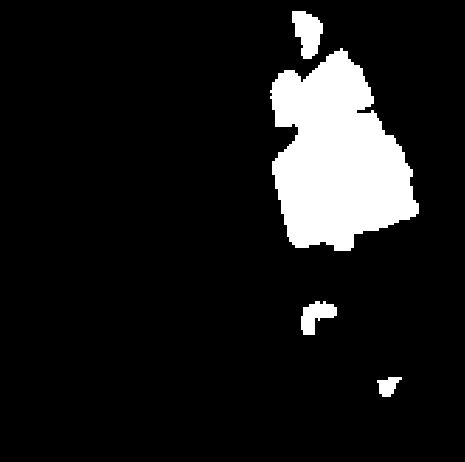}};
	\node at (5.20+\shift+0.1*3, 1+0.15*3, 0.75*3) {\includegraphics[width=0.16\textwidth]{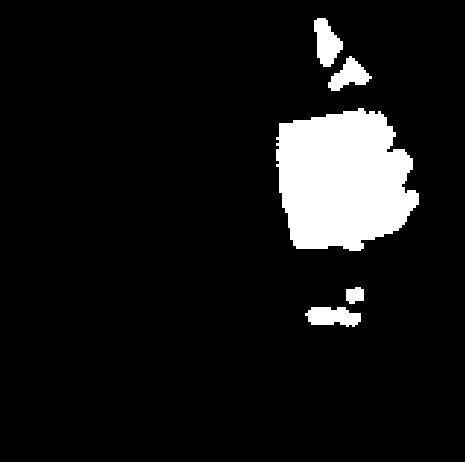}};

	\node at (7.2+\shift+0.1*3+\dx, 6.2+0.15*3, 0.75*3) {\scriptsize{Deconv CBF}}; 
	\node at (7.2+\shift+0.1*3+\dx, 5+0.15*3, 0.75*3) {\includegraphics[width=0.16\textwidth]{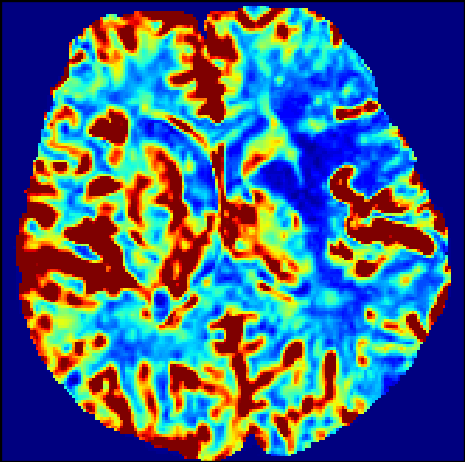}};
	\node at (7.2+\shift+0.1*3+\dx, 3+0.15*3, 0.75*3) {\includegraphics[width=0.16\textwidth]{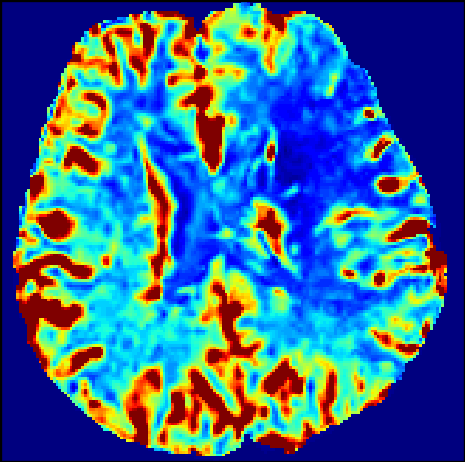}};
	\node at (7.2+\shift+0.1*3+\dx, 1+0.15*3, 0.75*3) {\includegraphics[width=0.16\textwidth]{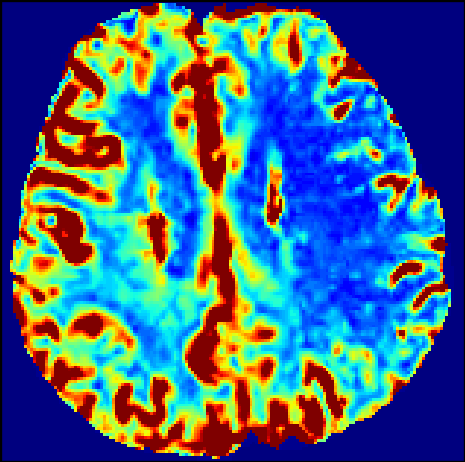}};

	\node at (9.2+\shift+0.1*3+\dx, 6.2+0.15*3, 0.75*3) {\scriptsize{HemoPIC CBF}}; 
	\node at (9.2+\shift+0.1*3+\dx, 5+0.15*3, 0.75*3) {\includegraphics[width=0.16\textwidth]{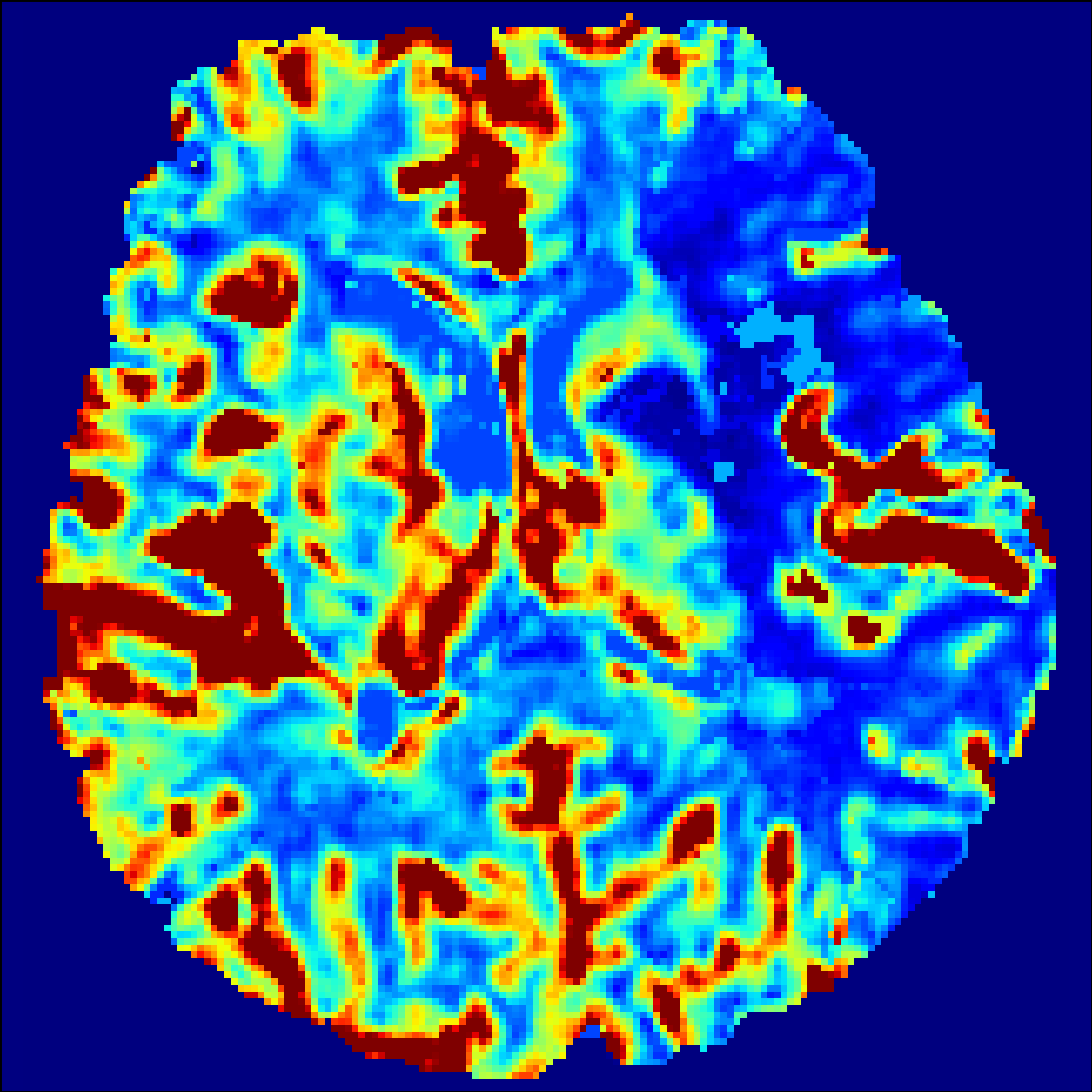}};
	\node at (9.2+\shift+0.1*3+\dx, 3+0.15*3, 0.75*3) {\includegraphics[width=0.16\textwidth]{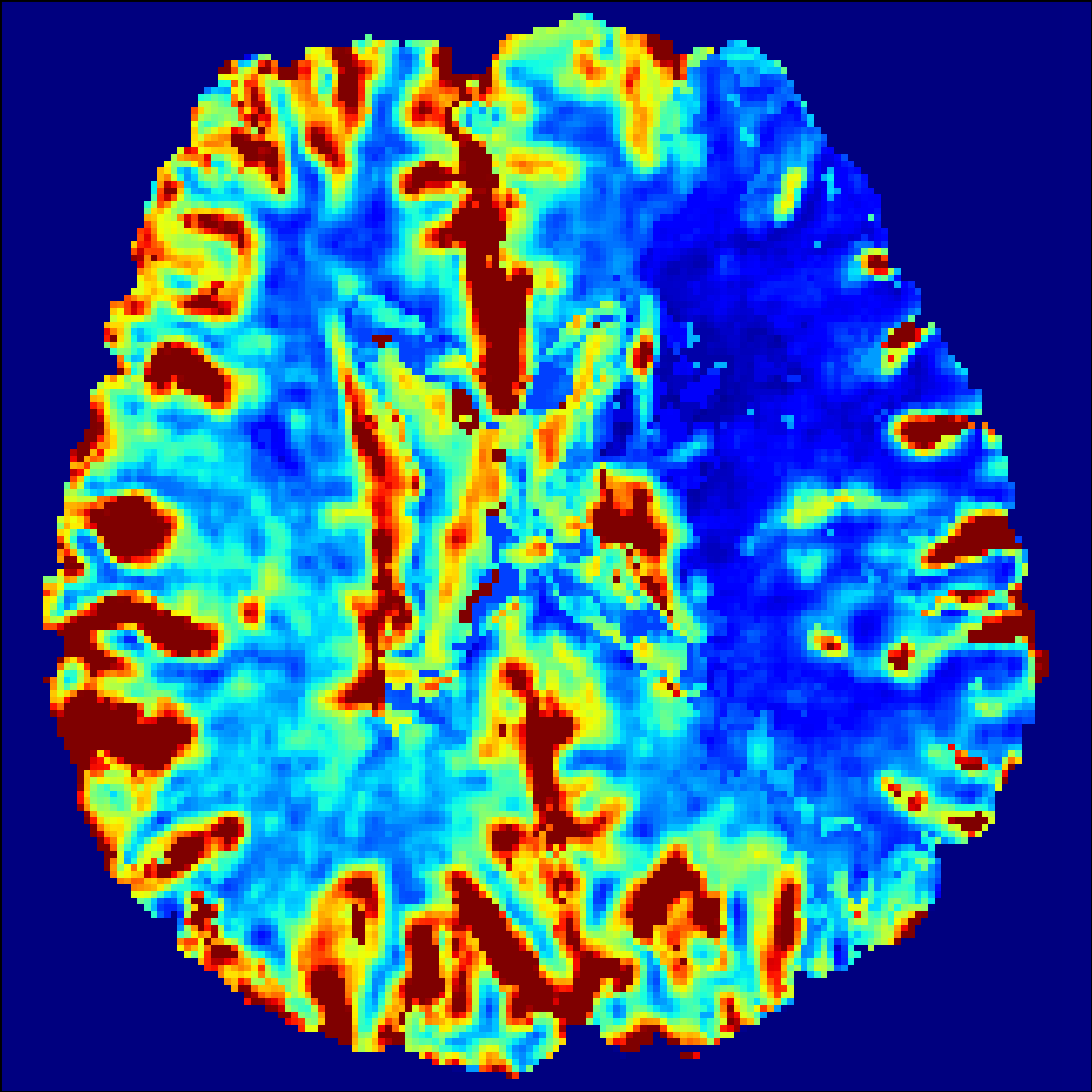}};
	\node at (9.2+\shift+0.1*3+\dx, 1+0.15*3, 0.75*3) {\includegraphics[width=0.16\textwidth]{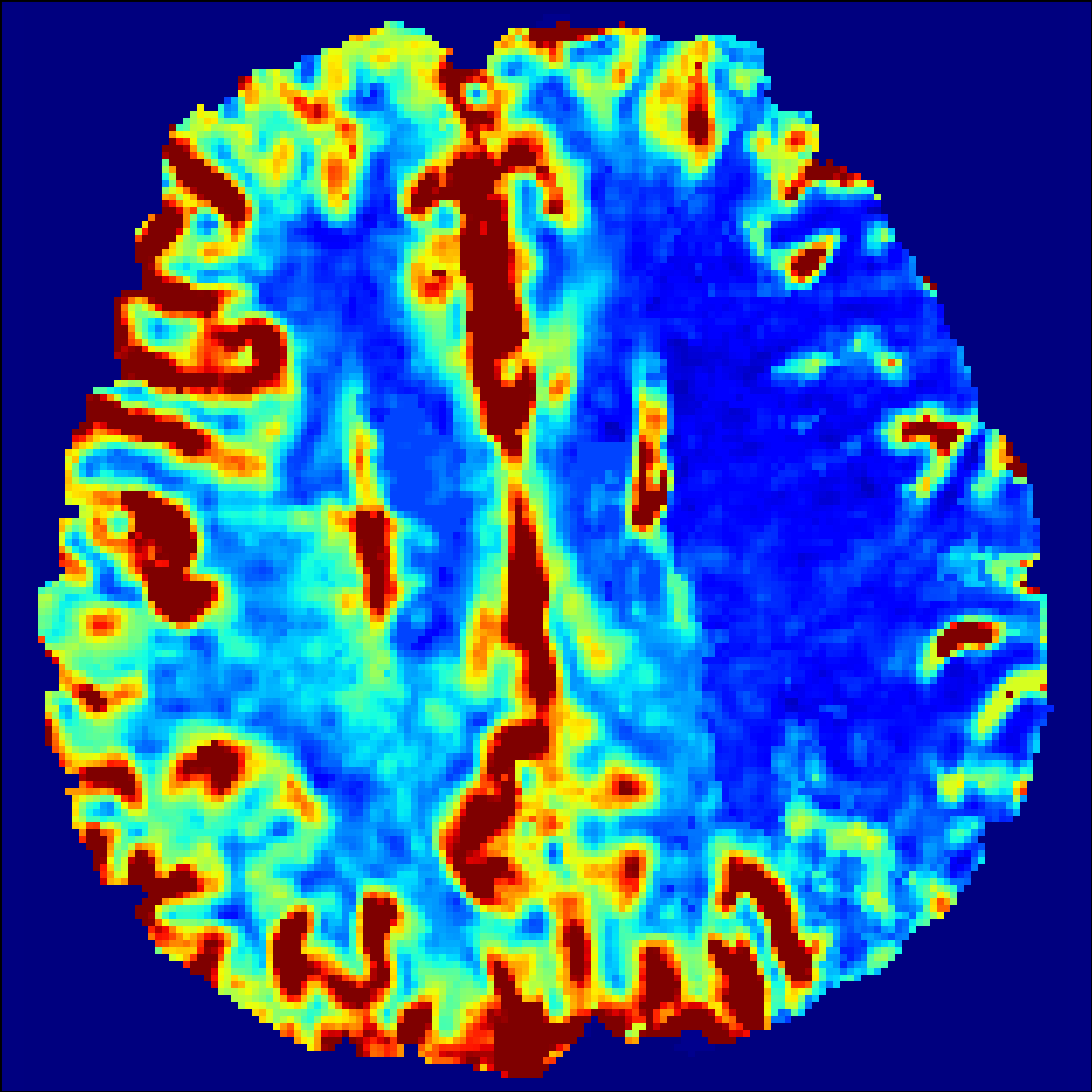}}; 
    
	\node at (8.20+\shift+0.1*3+\dx, -0.15+0.15*3, 0.75*3) {\includegraphics[width=0.13\textwidth]{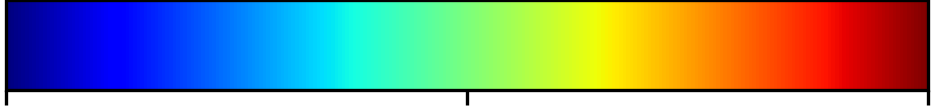}};
	\node at (8.20+\shift+0.1*3-0.78+\dx, -0.4+0.15*3, 0.75*3) {\scriptsize{0}};
	\node at (8.20+\shift+0.1*3+\dx, -0.4+0.15*3, 0.75*3) {\scriptsize{35}};
	\node at (8.20+\shift+0.1*3+0.78+\dx, -0.4+0.15*3, 0.75*3) {\scriptsize{70}};

	\node at (11.2+\shift+0.1*3+2*\dx, 6.2+0.15*3, 0.75*3) {\scriptsize{Deconv CBV}}; 
	\node at (11.2+\shift+0.1*3+2*\dx, 5+0.15*3, 0.75*3) {\includegraphics[width=0.16\textwidth]{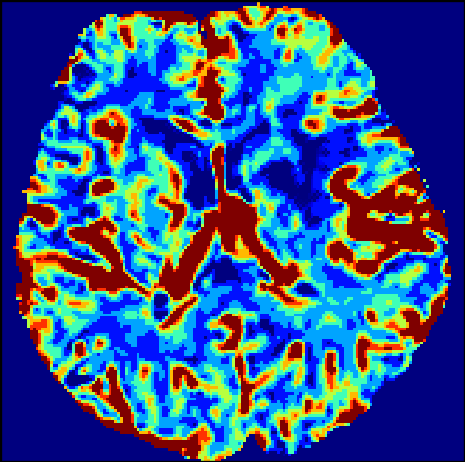}};
	\node at (11.2+\shift+0.1*3+2*\dx, 3+0.15*3, 0.75*3) {\includegraphics[width=0.16\textwidth]{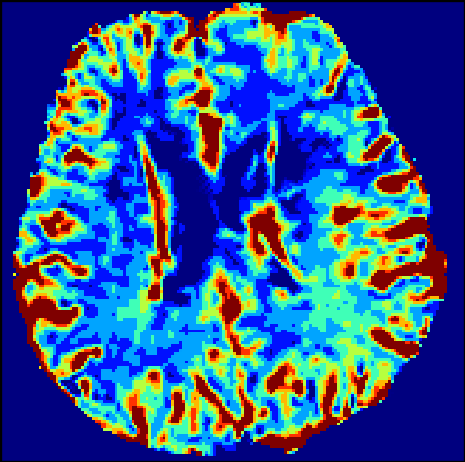}};
	\node at (11.2+\shift+0.1*3+2*\dx, 1+0.15*3, 0.75*3) {\includegraphics[width=0.16\textwidth]{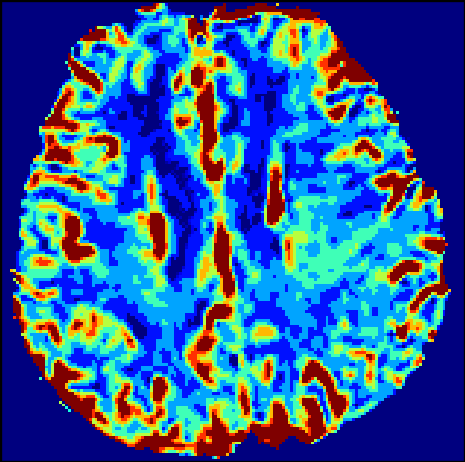}};
    
	\node at (13.2+\shift+0.1*3+2*\dx, 6.2+0.15*3, 0.75*3) {\scriptsize{HemoPIC CBV}}; 
	\node at (13.2+\shift+0.1*3+2*\dx, 5+0.15*3, 0.75*3) {\includegraphics[width=0.16\textwidth]{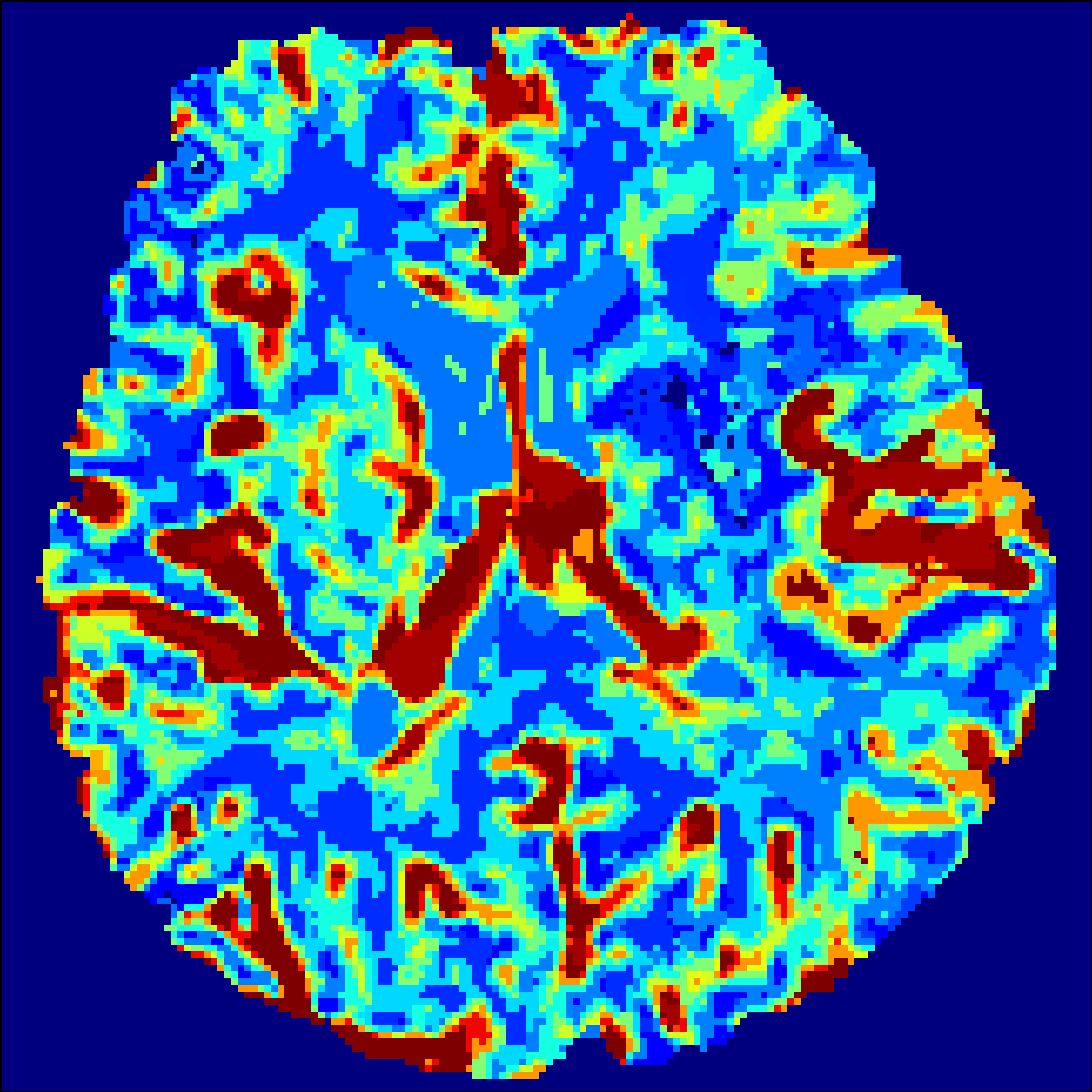}};
	\node at (13.2+\shift+0.1*3+2*\dx, 3+0.15*3, 0.75*3) {\includegraphics[width=0.16\textwidth]{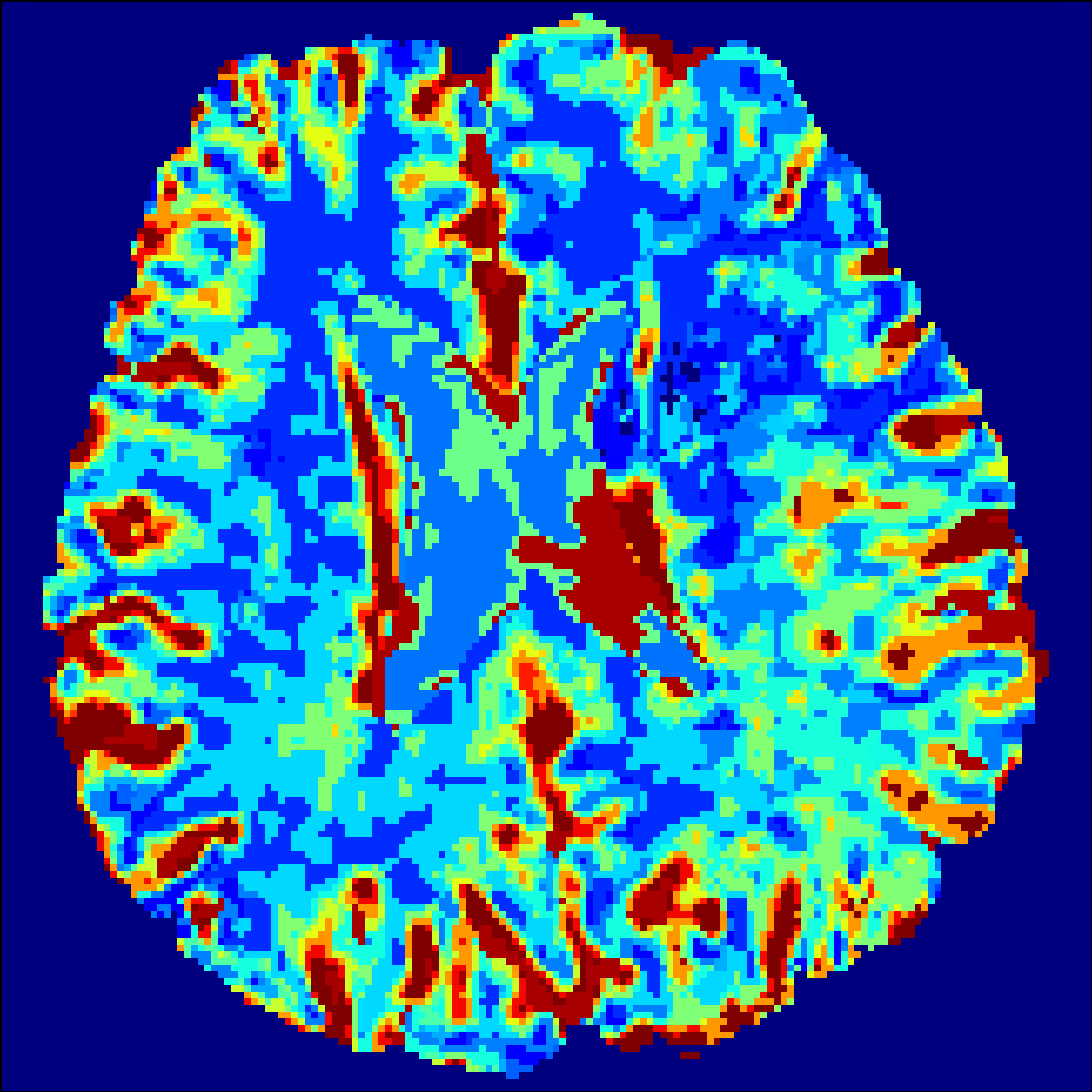}};
	\node at (13.2+\shift+0.1*3+2*\dx, 1+0.15*3, 0.75*3) {\includegraphics[width=0.16\textwidth]{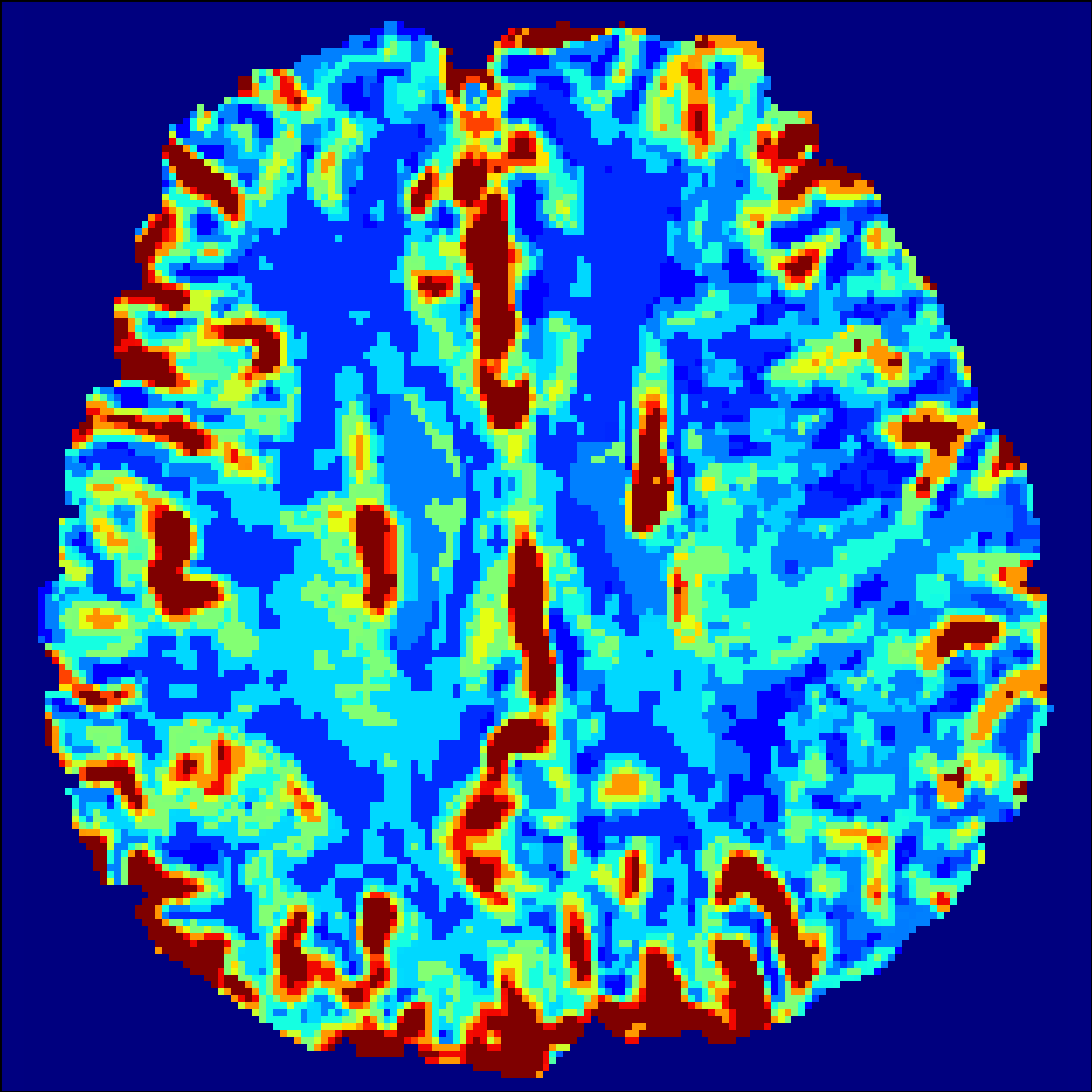}};
    
	\node at (12.20+\shift+0.1*3+2*\dx, -0.15+0.15*3, 0.75*3) {\includegraphics[width=0.13\textwidth]{figs/Fig2_Components/bar.png}}; 
	\node at (12.20+\shift+0.1*3-0.78+2*\dx, -0.4+0.15*3, 0.75*3) {\scriptsize{0}};
	\node at (12.20+\shift+0.1*3+2*\dx, -0.4+0.15*3, 0.75*3) {\scriptsize{3.5}};
	\node at (12.20+\shift+0.1*3+0.78+2*\dx, -0.4+0.15*3, 0.75*3) {\scriptsize{7}};

	\node at (15.2+\shift+0.1*3+3*\dx, 6.2+0.15*3, 0.75*3) {\scriptsize{Deconv MTT}}; 
	\node at (15.2+\shift+0.1*3+3*\dx, 5+0.15*3, 0.75*3) {\includegraphics[width=0.16\textwidth]{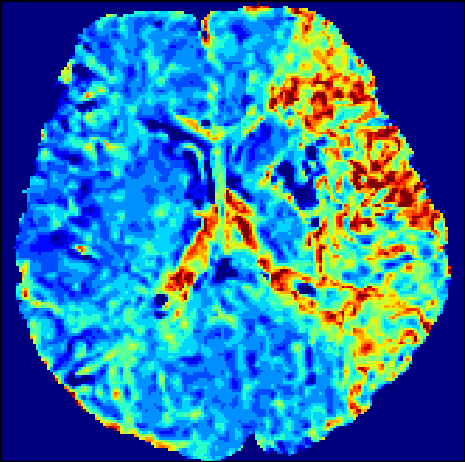}};
	\node at (15.2+\shift+0.1*3+3*\dx, 3+0.15*3, 0.75*3) {\includegraphics[width=0.16\textwidth]{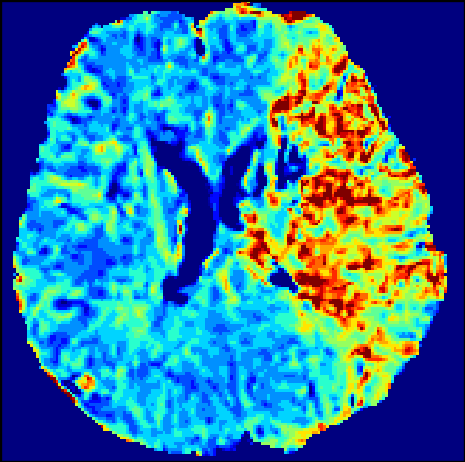}};
	\node at (15.2+\shift+0.1*3+3*\dx, 1+0.15*3, 0.75*3) {\includegraphics[width=0.16\textwidth]{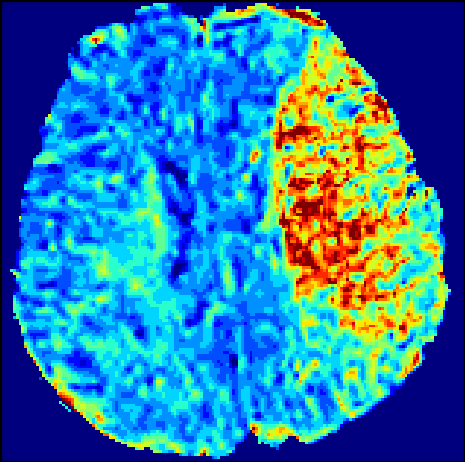}};
    
	\node at (17.2+\shift+0.1*3+3*\dx, 6.2+0.15*3, 0.75*3) {\scriptsize{HemoPIC MTT}}; 
	\node at (17.2+\shift+0.1*3+3*\dx, 5+0.15*3, 0.75*3) {\includegraphics[width=0.16\textwidth]{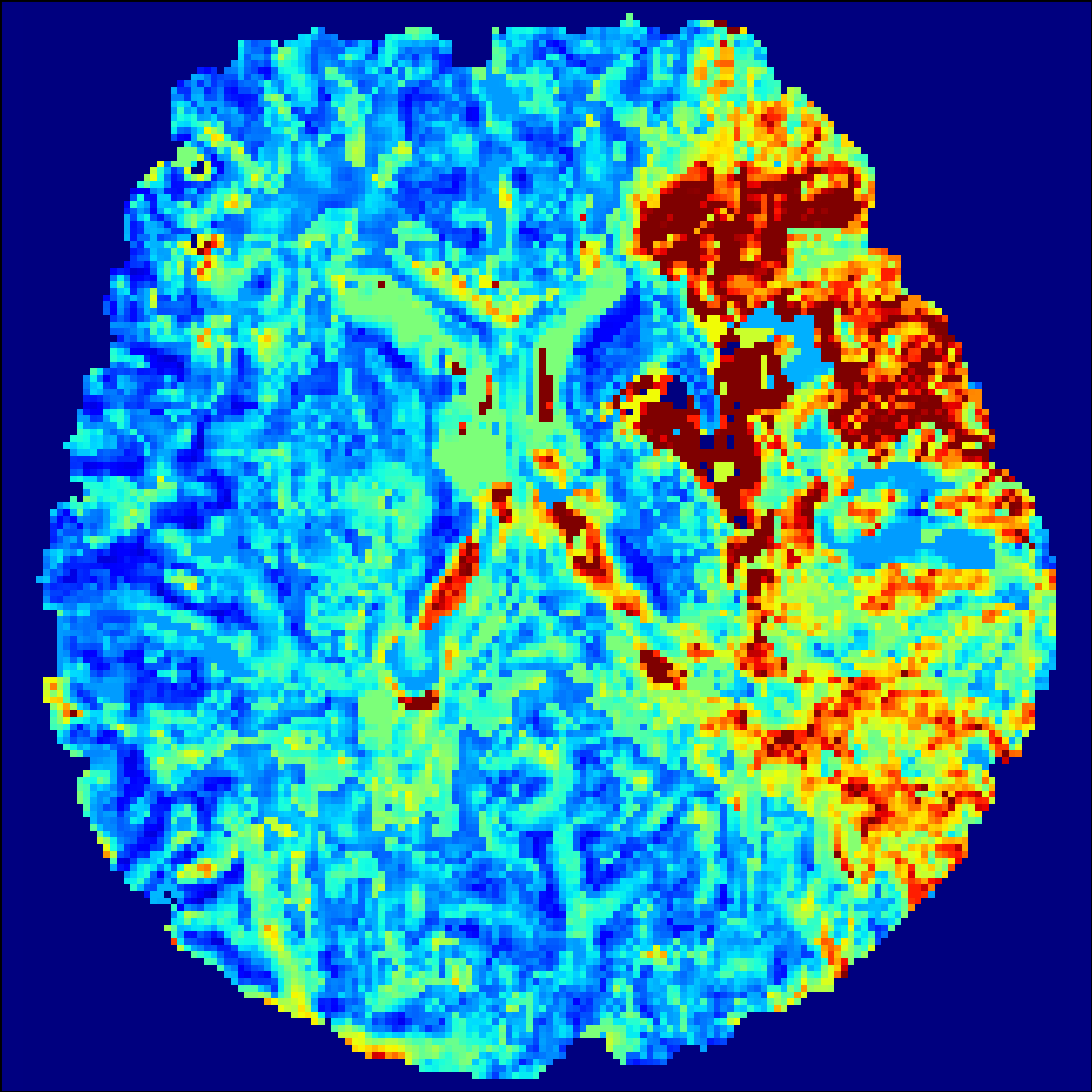}};
	\node at (17.2+\shift+0.1*3+3*\dx, 3+0.15*3, 0.75*3) {\includegraphics[width=0.16\textwidth]{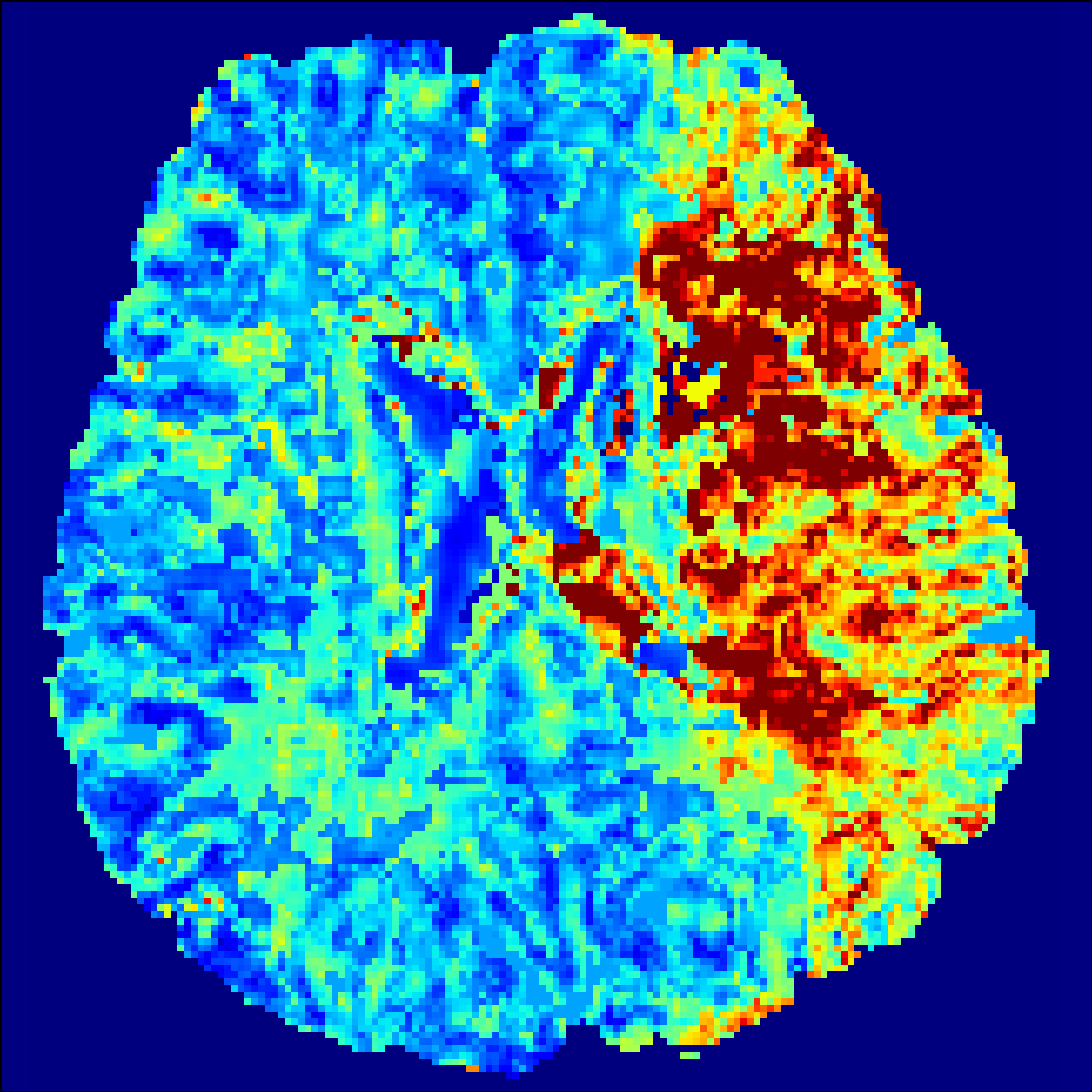}};
	\node at (17.2+\shift+0.1*3+3*\dx, 1+0.15*3, 0.75*3) {\includegraphics[width=0.16\textwidth]{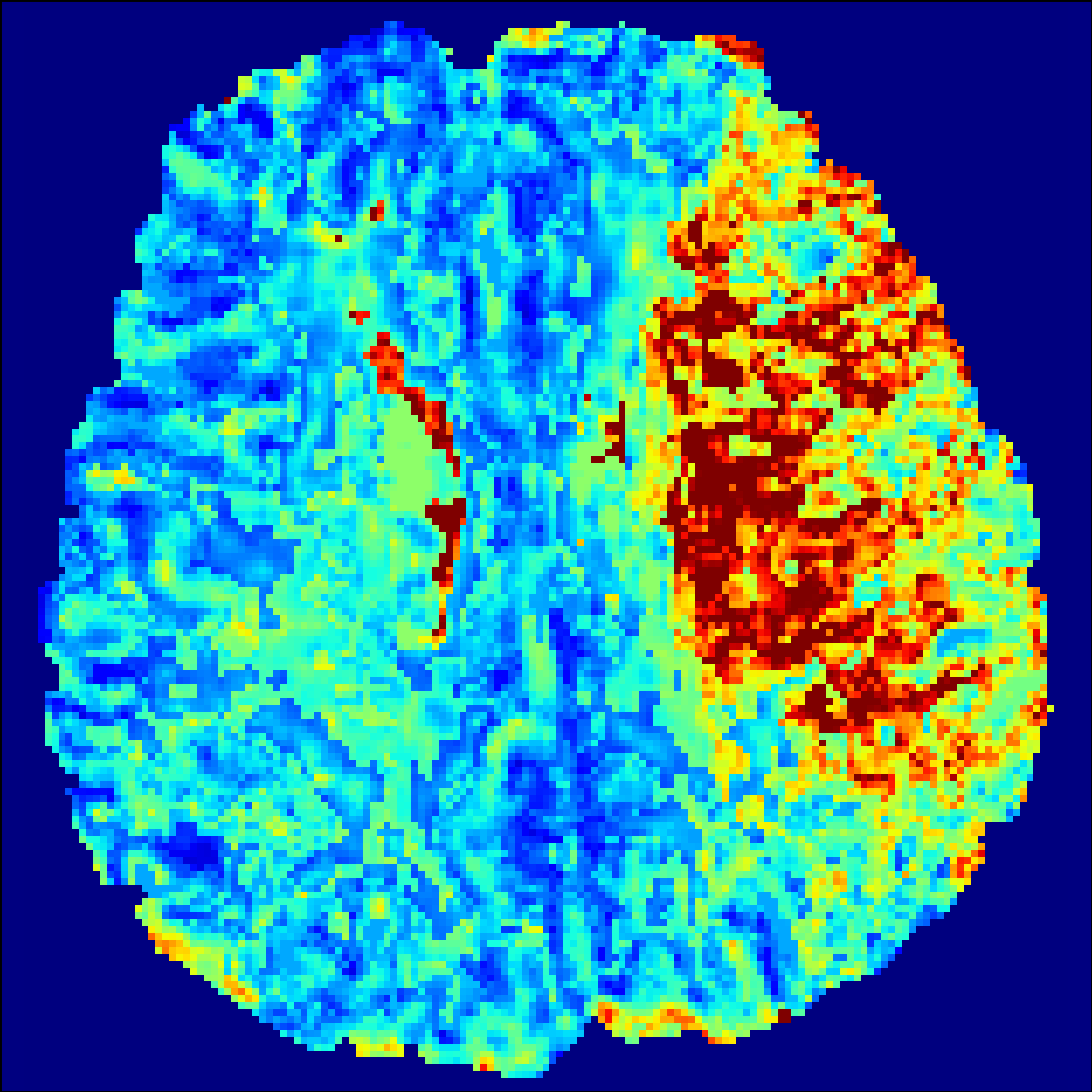}};
    
	\node at (16.20+\shift+0.1*3+3*\dx, -0.15+0.15*3, 0.75*3) {\includegraphics[width=0.13\textwidth]{figs/Fig2_Components/bar.png}};
	\node at (16.20+\shift+0.1*3-0.78+3*\dx, -0.4+0.15*3, 0.75*3) {\scriptsize{0}};
	\node at (16.20+\shift+0.1*3+3*\dx, -0.4+0.15*3, 0.75*3) {\scriptsize{7.5}};
	\node at (16.20+\shift+0.1*3+0.78+3*\dx, -0.4+0.15*3, 0.75*3) {\scriptsize{15}};


	\end{tikzpicture}
	}   
\caption{Qualitative comparison of perfusion summary maps from deconvolution-based models~\cite{winzeck2018isles} and HemoPIC, on an example case, with identical scales for each parameter.
} 

	 \label{fig:p22_example} 
\end{figure}

\begin{figure}[th]
  \centering
  \includegraphics[height=0.23\textheight]{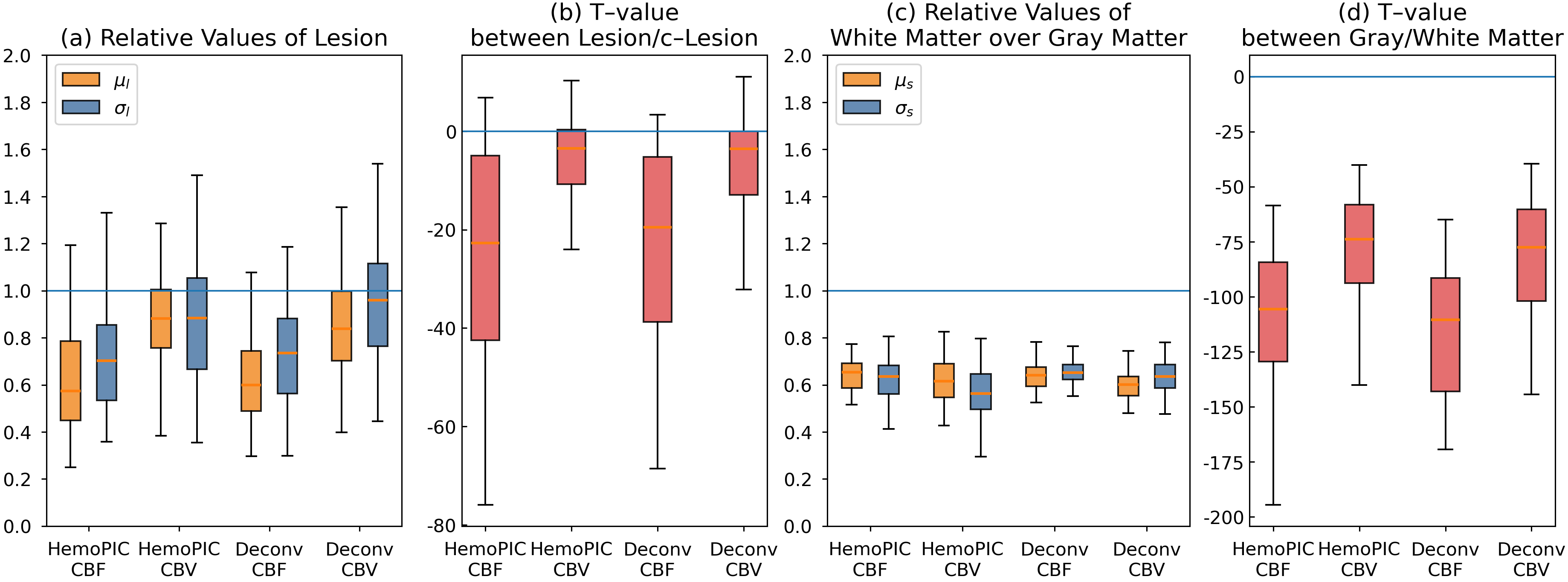}
  \caption{Box plots comparing HemoPIC and deconvolution-based (Deconv) maps (lower is better).
  Relative mean and std for lesion vs. c-lesion (a), and WM vs. GM (c). Corresponding unpaired t-value for lesion vs. c-lesion (b), and WM vs. GM (d). 
  }
  \label{fig:GMWM_Lesion}
\end{figure}



\begin{figure}[th]
  \centering
  \includegraphics[height=0.16\textheight]{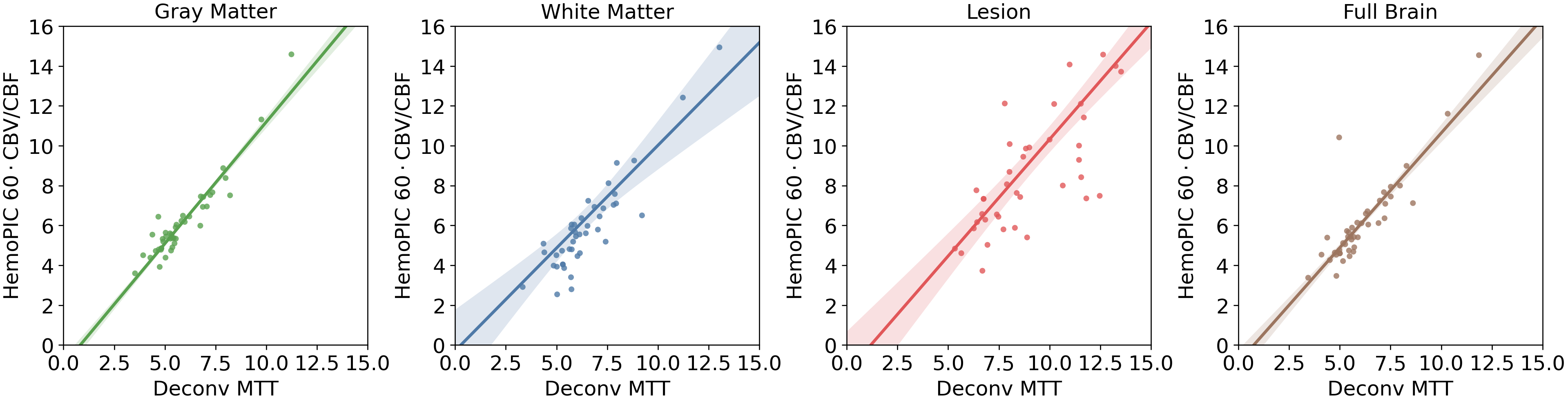}
  \caption{Regional central volume theorem consistency verification. Dots: patient-level regional means. Sines: region-wise OLS fits across $44$ patients with $\pm$ standard error bands. Slopes: 1.22 (GM), 1.03 (WM), 1.17 (Lesion), and 1.15 (Full Brain). 
  }
  \label{fig:CVT}
\end{figure}


\begin{figure}[th]
  \centering
  \includegraphics[height=0.19\textheight]{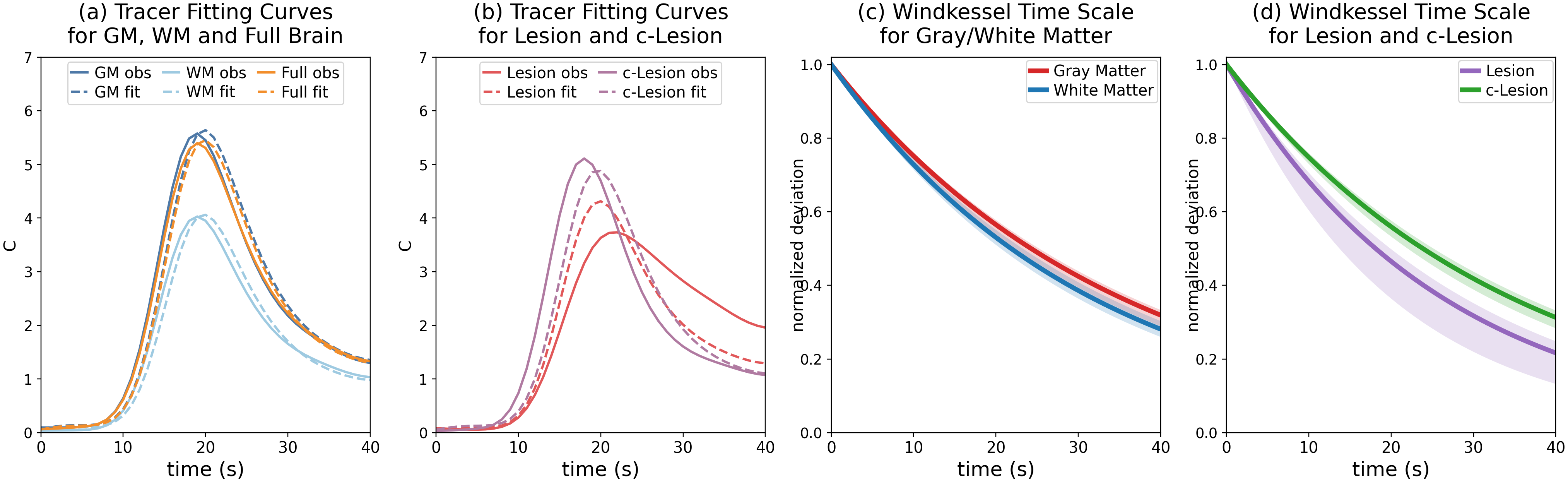}
  \caption{
  (a)-(b) Cohort-mean observed (solid) and reconstructed (dashed) tracer concentration time curves. (c)-(d) Windkessel time scales with voxel-weighted median (thick) and 40th-60th percentile band over the four regions across all 44 patients. 
  }
  \label{fig:ctc_fit}
\end{figure}

\subsection{Comparisons with AIF-based, Deconvolution Approaches}
\label{subsec:comp_results}
We first compare the perfusion summary maps (CBF, CBV, MTT) from conventional deconvolution-based models and HemoPIC.
We report regional CBF and CBV on the same numerical scale used conventionally in deconvolution-based models. MTT is reported in seconds.
We set the regional mean CBF in $\Omega_r$ to the mean over the last $20\%$ of the estimated blood outflow $\hat f^r_{\mathrm{out}}(t)$, then modulate it voxelwise within $\Omega_r$ using each voxel’s tracer curve to capture within-cluster heterogeneity, excluding non-parenchymal regions.
Similarly, we set the regional mean CBV in $\Omega_r$ to the estimated effective volume $V_r$, and distribute it voxelwise within $\Omega_r$ using the relative pattern of the data CBV map after clipping and mean normalization. MTT is computed by the relation $\mathrm{MTT} = 60 \cdot \mathrm{CBV}/\mathrm{CBF}$.

\paragraph{\textbf{Perfusion Summary Maps.}}
\cref{fig:p22_example} shows a qualitative comparison of the resulting perfusion summary maps from conventional, deconvolution-based approaches~\cite{winzeck2018isles} and HemoPIC, using identical scales for each parameter. 
Our maps not only match the scale and spatial pattern of the deconvolution-based maps, but also exhibit a stronger contrast for lesion delineation. 
Notably, these deconvolution pipelines all require manual AIF selection, making the resulting perfusion maps sensitive to algorithmic choices and potentially influencing downstream clinical decisions~\cite{kudo2010differences}. 
\cref{fig:GMWM_Lesion} further reports the relative mean and std deviation, with the unpaired t-statistic between voxel values in each region pair. Following~\cite{liu2020piano,liu2021perfusion,liu2021discovering}, we define $\mu_s$ and $\sigma_s$ as the WM to GM ratios of the regional mean, std, and define $\mu_l$ and $\sigma_l$ as the lesion to contralateral region (c-lesion, defined by mirroring the lesion across the cerebral hemispheric midline onto the contralateral unaffected side) ratios of the regional mean, std. 
Lower values indicate better performance for all comparisons. 
Importantly, $\mu_s < 1$ reflects higher relative mean CBF in GM than WM, consistent with cerebral perfusion physiology in healthy brains~\cite{fantini2016cerebral}.
Further, $\mu_l < 1$ reflects reduced lesion perfusion relative to c-lesion, consistent with acute ischemic lesion hypoperfusion~\cite{lu2018journal}.

\paragraph{\textbf{Regional Central Volume Theorem Consistency.}}
We further assess the consistency of the regional central volume theorem between HemoPIC and clinical perfusion summary maps. \cref{fig:CVT} compares the deconvolution-derived regional mean MTT against the regional mean $60 \cdot \mathrm{CBV}/\mathrm{CBF}$ from HemoPIC across all patients, in GM, WM, Lesion, and Full Brain. 
The scatter is approximately linear with slopes near one across regions, indicating quantitative agreement with the clinical MTT scale and the indicator dilution identity $\mathrm{MTT} = \mathrm{CBV}/\mathrm{CBF}$~\cite{meier1954theory,zierler2000indicator}.



\subsection{Unique Capabilities of HemoPIC}
\label{subsec:hemopic_results}

\paragraph{\textbf{Full Tracer Passage Reconstruction}.}
\cref{fig:ctc_fit} compares the observed regional mean concentration time curves $\bar{C}_r(t)$ and the reconstructed curves $\hat{C}_r(t)$ for GM, WM, lesion, the contralateral region
of the lesion (c-leion), and full brain. 
Quantitative reconstruction errors are summarized in \cref{table:ctc}. Importantly, prior tracer dynamics estimation models~\cite{liu2020piano,liu2021perfusion,liu2021discovering} can only estimate the outflow phase, i.e., time points following the global time-to-peak. 
In contrast, HemoPIC is the first approach capable of reconstructing the entire tracer dynamics, explicitly estimating inflow-dependent hemodynamic states that capture both inflow and outflow throughout the full tracer passage measured by perfusion imaging.


\paragraph{\textbf{Windkessel Dynamics.}}

We further report HemoPIC digital twin parameters that govern the regional Windkessel outflow dynamics through their product $\tau^r = \theta_R^r\theta_C^r$, which defines the characteristic relaxation time scale of \eqref{eqn:windkessel}. Assuming $f_{\mathrm{in}}^r(t) \approx F^r$, deviations of $f_{\mathrm{out}}^r(t)$ decay as $\exp(-t/\tau^r)$. We therefore report the normalized decay curve.
\cref{fig:ctc_fit} pools ROI-level $\tau^r$ across patients within each tissue group using voxel-wise weights, showing the decay curve induced by the weighted median (thick line) and the 40th–60th percentile range (shaded band). GM and WM exhibit similar relaxation (medians 13.95 s and 12.56 s). Lesions relax faster than contralateral tissue (medians 10.41 s vs. 13.71 s).

\section{Conclusion}
\label{sec:con}


We propose HemoPIC, a physics-informed digital twin for cerebral perfusion that couples hemodynamic modeling with tracer transport to infer patient-specific flow and volume parameters, directly from routine perfusion imaging. Beyond reconstructing full inflow–outflow tracer dynamics, HemoPIC generates clinically actionable perfusion maps without requiring manual AIF selection. To our knowledge, this is the first approach to link tracer transport modeling with clinically standard perfusion indices, enabling physics-grounded, fully automated stroke assessment and simulation-driven clinical decision support.

\clearpage

\bibliographystyle{splncs04}
\bibliography{ref}

\end{document}